\documentclass[lettersize,journal]{IEEEtran}
\usepackage{amsmath,amsfonts}
\usepackage{algorithmic}
\usepackage{algorithm}
\usepackage{amsthm}
\usepackage{array}
\usepackage{multirow}
\usepackage[caption=false,font=normalsize,labelfont=sf,textfont=sf]{subfig}
\usepackage{textcomp}
\usepackage{stfloats}
\usepackage{url}
\usepackage{verbatim}
\usepackage{graphicx}
\usepackage{cite}
\begin{document}

\title{A Piecewise-Linear Approximation-based Energy-Efficient Error-Optimized Unsigned Square Rooter for Accuracy-Critical Applications}

\author{Prateek Goyal,~\IEEEmembership{Student Member,~IEEE,} Sujit Kumar Sahoo,~\IEEEmembership{Senior Member,~IEEE}
\thanks{The authors are with the School of Electrical Sciences, Indian Institute of Technology Goa, Ponda-403401, Goa, India (e-mail: prateek22242202@iitgoa.ac.in; sujit@iitgoa.ac.in).}}

\maketitle

\begin{abstract}
 Approximate computing improves energy efficiency in error-resilient applications, but square root units remain challenging due to the trade-off between hardware cost and computational accuracy. This paper presents an energy-efficient, error-optimized, piecewise-linear approximation–based unsigned square rooter (EOSQR) for $2n$-bit inputs that achieves high accuracy with low hardware complexity, using only simple arithmetic and shift operations. The EOSQR design is implemented in Verilog-HDL and evaluated on a 16-bit benchmark synthesized on an Artix-$7$ FPGA. Compared to representative state-of-the-art approximate square rooters, EOSQR achieves the lowest error among accuracy-critical designs while delivering $61.91\%$ resource savings, $77.54\%$ power savings, and $53.11\%$ latency reduction relative to a precise restoring array-based square rooter. To enable holistic evaluation, a Composite Efficiency Metric (CEM) is introduced to jointly capture accuracy and energy efficiency. EOSQR is further validated on representative image-processing and machine-learning workloads, including Sobel edge detection, K-means color quantization, and K-nearest-neighbor (KNN) classification. Experimental results demonstrate that EOSQR achieves high computational accuracy with a superior CEM-based accuracy–hardware efficiency trade-off while maintaining visual quality and classification performance, making it well-suited for real-time edge-embedded systems.
 
\end{abstract}


\section{Introduction}
\IEEEPARstart{D}{igital} arithmetic forms the foundation of modern computing systems, supporting a wide range of applications from signal and image processing to graphics and machine learning accelerators \cite{AC1}. As system performance is increasingly constrained by power, latency, and hardware resources, approximate computing has emerged as an effective paradigm that exploits the inherent error resilience of many contemporary workloads. Within this framework, approximate arithmetic circuits, including adders \cite{AD1}\cite{AD2}\cite{AD_EOHEAA}, multipliers \cite{ML1}\cite{ML2}\cite{ML3}, dividers \cite{DI1}\cite{DI2}\cite{DI3}, and square root units \cite{SQ1}\cite{SQ2}\cite{OLSR}\cite{SQ3}\cite{TSOSQR}, have been widely explored to reduce computational complexity and energy consumption. Among these operations, square root computation is particularly important, as it is widely used in Euclidean distance evaluation, normalization, and feature extraction in image processing and machine learning. However, square root operations remain inherently resource- and energy-intensive, often relying on iterative or complex hardware designs that incur significant area, power, and latency overheads. 

Approximate square rooters can deliver substantial efficiency improvements by introducing controlled inaccuracies, thereby creating a fundamental trade-off between computational accuracy and hardware efficiency. Achieving an appropriate balance between these objectives is crucial, particularly in applications where accuracy degradation cannot be readily tolerated \cite{AC3}. In applications such as medical imaging, financial computing, high-quality image enhancement, and machine learning algorithms, including K-means clustering and K-nearest neighbors, square root computation is essential for evaluating Euclidean distances used in clustering and classification. These domains require high computational accuracy to ensure reliability while maintaining energy efficiency \cite{AC5}.

Consequently, accuracy-aware and error-optimized approximate square root designs have emerged as a critical requirement for extending approximate computing to reliability-sensitive applications. While several existing approaches aim to achieve error-optimized operation, many of these solutions improve accuracy at the expense of increased hardware complexity, resulting in higher resource utilization, power consumption, and delay \cite{SQ1}\cite{SQ3}. This limits their effectiveness in energy-constrained and real-time systems, where both accuracy and efficiency must be carefully balanced. Therefore, there remains a strong need for square root designs that deliver near-exact computational accuracy while maintaining comparable hardware overhead and a favorable accuracy–efficiency trade-off. Addressing this challenge requires a critical review of existing unsigned square root computation techniques and their inherent design trade-offs, discussed next.

The Exact Restoring Array-based Square Rooter (ERAS) \cite{AC1} employs an iterative shift-and-subtract restoring array algorithm to compute the exact square root of an unsigned number, thereby serving as a standard baseline for accuracy and hardware evaluation. Building on this framework, an Approximate Square Rooter (AXSR3) is introduced in \cite{SQ1}, in which selected exact subtractor cells within the restoring array are replaced by approximate cells using a triangular replacement strategy to generate an approximate square root. Restoring array–based methods rely on iterative subtraction and slow convergence, making them inefficient for large word lengths. As a result, series expansion–based square root methods are preferred for better scalability and computational efficiency.

Series expansion–based square root methods, such as the Energy-Efficient Logarithmic Square Rooter with Error Compensation (LESQ-EC) \cite{SQ2}, the Optimized Logarithmic Square Rooter (OLSR) \cite{OLSR}, and the Taylor Series–based Optimal Square Rooter (TSOSQR) \cite{TSOSQR}, approximate the square root by decomposing the input radicand into its nearest power-of-two component and a residual term, which is estimated using a truncated series expansion. While this strategy enables shift-and-add implementations with reduced hardware complexity and energy consumption, its reliance on low-order series approximations and MSB-dominant representations limits numerical accuracy. The error becomes more pronounced for inputs with larger residual components, underscoring the need for alternative square-root designs that enhance accuracy while preserving a balanced accuracy–efficiency trade-off.

The Modified Approximate Hybrid Square Rooter (MAHSQR) \cite{SQ3} is a hybrid architecture that integrates a logarithmic approximation scheme with a reduced-bit-width Exact Restoring Array Square Rooter (ERAS) to balance computational accuracy and hardware efficiency. It computes $\sqrt{M}$ by partitioning the input radicand ({M}) into its most significant and least significant segments and applying a first-order linear approximation. To improve accuracy, only a small portion of the least significant bits (LSBs) is approximated, while most of the input is processed using an exact restoring array unit. Although this enhances precision, the larger exact block increases hardware overhead, leading to higher area, critical path delay, and power–delay product (PDP), thereby limiting suitability for hardware-constrained applications.
\subsection{Critical Evaluation of Existing Limitations and Motivation for Enhancement}
Although several approximate square root designs have been proposed to reduce hardware cost, notable limitations arise when accuracy is prioritized. Restoring array-based designs, such as AXSR3 \cite{SQ1}, employ iterative subtract-and-restore operations within regular array structures, providing deterministic and relatively accurate computation; however, their topology incurs substantial hardware overhead, including high LUT utilization, increased switching activity, long critical paths, and poor scalability with increasing operand bit-width. Hybrid square rooters such as MAHSQR \cite{SQ3} attempt to balance accuracy and efficiency by computing the most significant bits using an Exact Restoring Array Square Rooter (ERAS) \cite{AC1} while approximating the remaining portion, but their reliance on restoring arrays preserves similar scalability bottlenecks and hardware costs. 

Although restoring array-based approaches improves accuracy, they incur higher area, delay, and power, limiting gains in power-delay product (PDP). 
In contrast, series-expansion–based square-root designs \cite{SQ2}\cite{OLSR}\cite{TSOSQR} significantly reduce hardware complexity by employing truncated approximations implemented via shift-and-add operations. However, they exhibit non-uniform error behavior across different approximation intervals, resulting in continuity mismatches at interval boundaries and noticeable deviations from the true square-root curve near transition points. Their limited numerical accuracy makes them less suitable for accuracy-critical 
applications requiring high accuracy with low hardware complexity. 

Consequently, existing square-root designs face a fundamental trade-off between hardware efficiency and computational accuracy. Hardware-efficient designs often sacrifice accuracy, whereas accuracy-preserving approaches incur significant area, delay, and power overhead.
This trade-off limits their applicability in accuracy-sensitive domains such as Euclidean distance–based clustering (e.g., K-means) and classification (e.g., K-nearest neighbors), as well as high-quality imaging tasks, where even small computational errors can propagate and degrade system-level performance. Therefore, there is a strong need for an error-optimized square rooter that achieves near-exact computational accuracy while preserving hardware efficiency, enabling a well-balanced accuracy–efficiency trade-off without incurring the scalability and energy penalties associated with conventional exact square root designs.

\subsection{Highlights of the Proposed Work}
This work presents an energy-efficient piecewise-linear approximation–based Error-Optimized Unsigned Square Rooter (EOSQR) that achieves high computational accuracy with reduced hardware cost and power-delay product (PDP) compared with existing designs of similar accuracy, while improving the overall 
accuracy–efficiency trade-off as quantified by the Composite Efficiency Metric (CEM). The major contributions of this work are summarized as follows:
\begin{itemize}
\item A novel piecewise-linear, error-optimized square-root approximation framework is proposed, using analytical optimization to achieve near-accurate results.
\item The proposed EOSQR design is implemented in Verilog-HDL and synthesized using Xilinx Vivado $2019.2$. Its performance is evaluated using standard accuracy and hardware efficiency metrics, followed by a comprehensive graphical comparison.
\item A composite efficiency metric (CEM) framework is introduced to analyze the accuracy–efficiency trade-off, correlating error metrics with PDP and demonstrating the superior balance of EOSQR over state-of-the-art designs.
\item The practical applicability of EOSQR is demonstrated through image processing and machine learning tasks, including Sobel edge detection, K-means clustering, and KNN classification.
\end{itemize}

The remainder of this paper is organized as follows. Section $2$ presents the proposed piecewise-linear, error-optimized square-root approximation framework, including its error optimization and correction mechanisms. Section $3$ presents graphical analysis, accuracy evaluation, and a comprehensive comparison with existing designs, including a CEM-based assessment of accuracy-efficiency trade-offs. Section $4$ demonstrates the effectiveness of the proposed design through image processing and machine learning applications with visual results. Section $5$ concludes the paper.
\section{Piecewise-Linear Approximation via Error Optimization for Proposed Square Rooter}

Let \(M\) be a \(2n\)-bit unsigned integer that can be expressed as
\[
M = 2^r + y,
\]
where \(2^r\) denotes the largest power of two not exceeding \(M\), \(r\) is the position of the leading one in the binary representation of \(M\), and \(y\) represents the residual remainder satisfying \(0 \le y < 2^r\). The square root function $\sqrt{M}$ can be approximated in the interval $ y \in [0, 2^r - 1]$ using a first-order piecewise-linear approximation model:
\begin{align}
\sqrt{2^r + y} \approx f_r(y) = b_r + m_ry
\end{align}

To obtain optimal parameters \(b_r\) and \(m_r\), the approximation error is defined as follows:
\begin{align*}
    E_r(y) = \sqrt{2^r + y} - (b_r + m_ry)
\end{align*}

We want to find the values of $b_r$ and $m_r$ such that the linear approximation meets the function at the endpoints \(2^r\) and \(2^{r+1}\) of the piece of the real axis we are approximating. i.e.
\[f_r(y)=\sqrt{2^r+y}\implies E_r(y) = 0 \text{ at } y  = \{0, 2^r\}\]
By forcing this matching at the endpoints of the interval, we have ensured that the piecewise linear approximation maintains the continuity property of the square root function. 

\subsubsection*{\underline{{Conditioning at \(y = 0\)}}}
At the lower end point:
\begin{align}
f_r(0) = \sqrt{2^r} \implies \boxed{ b_r = 2^{r/2}}
\end{align}

\subsubsection*{\underline{{Conditioning at \(y = 2^r\)}}}
At the upper endpoint:
\begin{align*}
\implies b_r + m_r2^r= \sqrt{2^r + 2^r}\\
\text{Substituting \(b_r = 2^{r/2}\)}\\
\implies 2^{r/2} + m_r2^r =\sqrt{2^{r+1}} 
\end{align*}

Thus, the slope is obtained as:
\begin{align*}
m_r = \frac{\sqrt{2^{r+1}} - 2^{r/2}}{2^r}
\end{align*}

\subsubsection*{Final Expression}

The resulting linear approximation is:
\begin{align*}
f_r(y) = 2^{r/2} + 
\frac{\sqrt{2^{r+1}} - 2^{r/2}}{2^r}\, y
\end{align*}

The value of the slope \(m_r\):
\begin{align}
m_r = \frac{2^{(r+1)/2} - 2^{r/2}}{2^r} \implies \boxed{m_r= \frac{1}{2^{r/2}} \left(\sqrt{2} - 1\right)}
\end{align}
Accordingly, the optimized linear square-root approximation can be written as:
\begin{align}
\boxed{\sqrt{M} \approx f_r(y) = 2^{r/2} + \frac{y}{2^{r/2}}\left(\sqrt{2}-1\right)}
\label{eq:sa1}
\end{align}
over the interval \(M \in [2^r,\,2^{r+1}]\). Furthermore, since the slope decreases exponentially with increasing \(r\), the contribution of the residual term \(y\) progressively diminishes for larger input magnitudes. This behavior reflects the concave nature of the square-root function and justifies the use of a first-order linear approximation within each interval \(M \in [2^r,\,2^{r+1}]\).
\begin{figure}[ht!]
\centering
\includegraphics[scale=0.55]{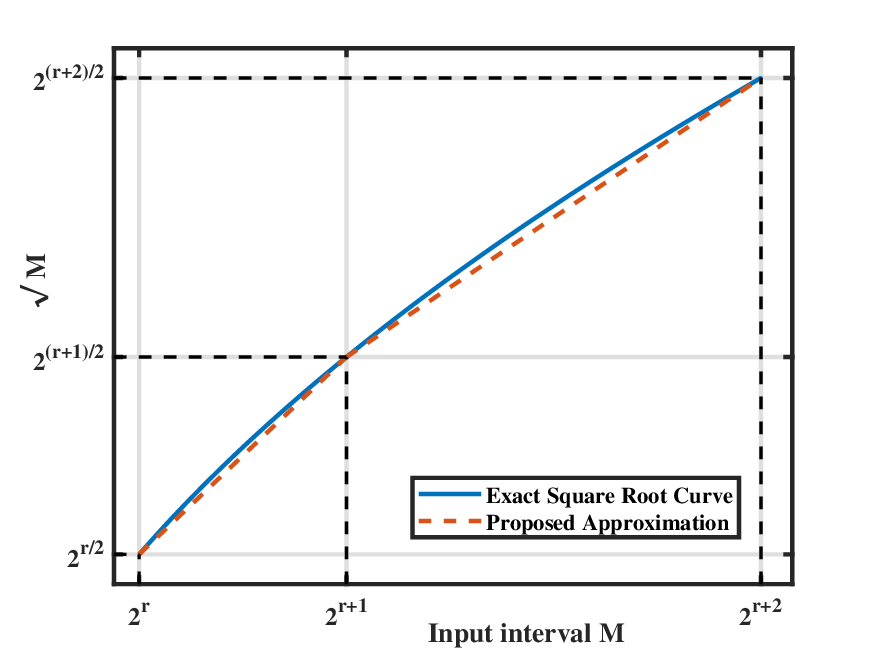}
 \caption{Geometric interpretation of the proposed endpoint-based approximation illustrating linear interpolation between successive square-root interval boundaries \(2^{r/2}\), \(2^{(r+1)/2}\), and \(2^{(r+2)/2}\), highlighting the scalability of the formulation for all \(2n\)-bit input operands.}
  \label{fig:mapping}
\end{figure}

As illustrated in Fig.~\ref{fig:mapping}, the square-root function over the interval \( M \in [2^r,\,2^{r+1}] \) maps to the corresponding output range 
\( \sqrt{M} \in [2^{r/2},\,2^{(r+1)/2}] \). Exploiting this relationship, 
a linear approximation is constructed by interpolating between these endpoints, enabling an efficient piecewise representation of the square-root function. 
The dotted projections in the figure illustrate this input-output interval mapping and highlight the geometric basis of the proposed approximation strategy employed. Since the parameters \(b_r=2^{r/2}\) and 
\(m_r=\frac{1}{2^{r/2}}(\sqrt{2}-1)\) are expressed explicitly in terms of the leading-one position \(r\), the formulation is independent of the absolute magnitude of the input operand and depends only on its normalized interval representation. Consequently, the same approximation structure can be uniformly applied across the entire range of the radicand, making the proposed model directly scalable to arbitrary word lengths. This interval-invariant formulation ensures that the approximation remains valid for all unsigned \(2n\)-bit numbers while maintaining a consistent balance between accuracy and energy efficiency.

The formulation in \eqref{eq:sa1} is mathematically valid for all values of \(r\); however, energy-efficient realization with comparable hardware utilization requires additional considerations, particularly for odd values of \(r\). In such cases, the term \(r/2\) introduces fractional exponents that cannot be directly implemented using shift-and-add operations alone. Moreover, the constant \((\sqrt{2}-1)\approx 0.4142\) must be expressed using an accurate shift-add representation to preserve multiplier-free computation for both even and odd values of \(r\). Therefore, suitable implementation strategies are adopted to accommodate these conditions while retaining the simplicity of the shift-based structure. These adjustments ensure that the proposed approximation maintains high energy efficiency and accuracy across all approximation intervals, yielding a compact, low-complexity design.

\subsubsection{\underline{Energy-Efficient Adjustments for Even \(r\)}}
For energy-efficient realization, the constant factor \((\sqrt{2}-1)\approx 0.4142\) in \eqref{eq:sa1} is approximated using dyadic fractions so that the multiplication with the residual term \(y\) can be implemented using simple shift-and-add operations. 
Accordingly, two practical dyadic approximations are considered for realizing the constant factor, corresponding to underestimation and overestimation cases, respectively:
\begin{equation*}
0.4142 \approx
\begin{cases}
\left(\dfrac{1}{2}-\dfrac{1}{8}\right)=0.375, & \text{Underestimation},\\[2mm]
\left(\dfrac{1}{2}-\dfrac{1}{16}\right)=0.4375, & \text{Overestimation}.
\end{cases}
\end{equation*}
Among the two cases, the overestimation approximation provides a closer representation; the approximation is selected accordingly. Therefore, for even values of $r$, the square-root approximation in \eqref{eq:sa1} can be expressed as follows: 
\begin{align}
\sqrt{M} \approx 2^{r/2} +
\frac{y}{2^{\frac{r}{2}}}
\Bigl(\frac{1}{2}-\frac{1}{16}\Bigr) 
\label{eq:sa}
\end{align}

\subsubsection*{\underline{Average Error Introduced Due to Implementation}}
The following is the error measure between the proposed linear approximation and the feasible hardware implementations in a given line segment of length $2^r$, for even values of $r$. 
\begin{align*}
E_{avg}
&= \frac{1}{2^r}
\sum_{y=0}^{2^r-1}
\left[
\frac{y}{2^{\frac{r}{2}}}
\left(\sqrt{2}-1\right)
-
\frac{y}{2^{\frac{r}{2}}}
\left(\frac{1}{2}-\frac{1}{16}\right)
\right] \nonumber \\[6pt]
&= \frac{1}{2^r}
\sum_{y=0}^{2^r-1}
\frac{y}{2^{\frac{r}{2}}}
\left(-0.02328\right) \nonumber \\[6pt]
&= \left(-0.02328\right)
\frac{1}{2^r}
\cdot
\frac{1}{2^{\frac{r}{2}}}
\cdot
\frac{2^r(2^r-1)}{2}  \nonumber \\[6pt]
&\approx 2^{\frac{r}{2}} \left( -0.01164 \right) 
\label{eq:avg_error}
\end{align*}

\subsubsection{\underline{Energy-Efficient Adjustments for Odd \(r\)}}
When \(r\) is odd, the term \(2^{r/2}\) introduces a fractional exponent that cannot be directly implemented using shift operations. To enable efficient realization, it is rewritten as
\begin{equation}
2^{\frac{r}{2}} = 2^{\frac{r-1}{2}} \cdot \sqrt{2} 
\label{eq:r_odd_decomp}
\end{equation}

Substituting \eqref{eq:r_odd_decomp} into \eqref{eq:sa1} gives
\begin{equation}
\sqrt{M} \approx
2^{\frac{r-1}{2}}\cdot\sqrt{2}
+
\frac{y}{2^{\frac{r-1}{2}}\cdot \sqrt{2}}
(\sqrt{2}-1)
\label{eq:r_odd_expanded}
\end{equation}

The above equation contains two terms that are approximated separately for efficient implementation.

\subsubsection*{\underline {First-Term Implementation}}

For energy efficiency and enabling realization using shifts and additions only, the factor \(\sqrt{2}\) in the first term of \eqref{eq:r_odd_expanded} can be approximated as:
\[
\sqrt{2} =1.41421 \approx \left(
1+\frac{1}{2}-\frac{1}{16}
\right) = 1.4375 =\lceil\sqrt{2} \rceil
\]

\begin{equation}
2^{\frac{r}{2}} =2^{\frac{r-1}{2}}\cdot\sqrt{2}
\approx
2^{\frac{r-1}{2}}
\left(
1+\frac{1}{2}-\frac{1}{16}
\right) = 2^{\frac{r-1}{2}}\lceil\sqrt{2} \rceil
\label{eq:first_term_approx}
\end{equation}
 where the factor $\lceil\sqrt{2} \rceil$ provides a hardware-efficient approximation for $\sqrt{2}$. When odd \(r\) is reformulated as \(r-1\) to eliminate fractional exponents, a level shift is introduced; this factor compensates for the shift, ensuring magnitude consistency and continuity while enabling efficient realization using adders and shifters.

\subsubsection*{First Term Implementation Error (due to $\lceil\sqrt{2} \rceil$)}
\begin{align}
E_1
&= 2^{\frac{r-1}{2}} \cdot 2^{\frac{1}{2}}
 - 2^{\frac{r-1}{2}}
\left( 1 + \frac{1}{2} - \frac{1}{16} \right) \nonumber \\
&\approx 2^{\frac{r}{2}} \left( -0.01646 \right)
\label{eq:first_term_error}
\end{align}

\subsubsection*{\underline {Second-Term Implementation}}

The second term in \eqref{eq:r_odd_expanded} can be simplified as
\[
\frac{y}{2^{\frac{r-1}{2}}\cdot \sqrt{2}}
(\sqrt{2}-1)
=
\frac{y}{2^{\frac{r+1}{2}}}
(2-\sqrt{2})
\]
Considering hardware and energy-efficient implementations, the constant multiplier is approximated using dyadic fractions, enabling multiplication via simple shift-and-add operations. Since $2-\sqrt{2} \approx 0.5857$, practical dyadic approximations are considered for the slope adjuster corresponding to the odd case of $r$. Accordingly, two candidate approximations are evaluated as:

\begin{equation*}
0.5857\approx
\begin{cases}
  \Bigl(\dfrac{1}{2}+\dfrac{1}{16}\Bigr)=0.5625, & \text{Underestimation},\\[2mm]
  \Bigl(\dfrac{1}{2}+\dfrac{1}{8}\Bigr)=0.6250 , & \text{Overestimation}.
  
\end{cases}
\end{equation*}

Since the first-term approximation slightly overestimates the result, as indicated in \eqref{eq:first_term_error}, the underestimation is selected for the second term to compensate for the overall approximation error. Hence, the second term is expressed as

\begin{equation}
\frac{y}{2^{\frac{r-1}{2}}\cdot \sqrt{2}}
(\sqrt{2}-1)
\approx
\frac{y}{2^{\frac{r+1}{2}}}
\left(\frac{1}{2}+\frac{1}{16}\right)
\label{eq:second_term_final}
\end{equation}
\subsubsection*{\underline{Average Error in Second Term Due to Implementation}}

\begin{align}
E_2
&= \frac{1}{2^r}
\sum_{y=0}^{2^r-1}
\left[
\frac{y}{2^{\frac{r}{2}}}
\left(\sqrt{2}-1\right)
-
\frac{y}{2^{\frac{r+1}{2}}}
\left(\frac{1}{2}+\frac{1}{16}\right)
\right] \nonumber \\[6pt]
&= \frac{1}{2^r}
\sum_{y=0}^{2^r-1}
\frac{y}{2^{\frac{r}{2}}}
\left(0.01647\right) \nonumber \\[6pt]
&= \left(0.01647\right)
\frac{1}{2^r}
\cdot
\frac{1}{2^{\frac{r}{2}}}
\cdot
\frac{2^r(2^r-1)}{2}  \nonumber \\[6pt]
&\approx 2^{\frac{r}{2}} \left( +0.00823 \right) 
\label{eq:avg_error_clean}
\end{align}
The errors contributed by the first and second terms in \eqref{eq:first_term_error} and \eqref{eq:avg_error_clean} largely neutralize each other, yielding an overall average error in the range $[2^r, 2^{r+1})$ in implementation:

\[
{E_{avg}}= {E1} + {E2}= 2^{\frac{r}{2}} \left( -0.00823 \right)
\]

Final approximation for Odd \(r\) by \eqref{eq:first_term_approx} and \eqref{eq:second_term_final} will follow:
\begin{equation}
\sqrt{M} \approx
2^{\frac{r-1}{2}}(\lceil\sqrt{2} \rceil)
+
\frac{y}{2^{\frac{r+1}{2}}}
\Bigl(\frac{1}{2}+\frac{1}{16}\Bigr) 
\label{eq:final_odd}
\end{equation}

\subsubsection{\underline{Overall Energy-Efficient Approximation}}
Using the derived approximations from \eqref{eq:sa} and \eqref{eq:final_odd}, the overall square-root computation can be expressed as
\begin{equation}
    \sqrt{M} \approx \begin{cases}
2^{r/2}
+
\frac{y}{2^{\frac{r}{2}}}\, \Bigl(\frac{1}{2}-\frac{1}{16}\Bigr), & \text{for }\textbf{even } r,\\ \\
2^{\frac{r-1}{2}}(\lceil\sqrt{2} \rceil)
+
\frac{y}{2^{\frac{r+1}{2}}}
\Bigl(\frac{1}{2}+\frac{1}{16}\Bigr), & \text{for } \textbf{odd } r.
\end{cases}
\label{eq:final}
\end{equation}

The approximation in \eqref{eq:final} decomposes the square-root operation into a dominant power-of-two base term and a residue-dependent correction component. The base term \(2^{r/2}\) (for even \(r\)) or \(2^{\frac{r-1}{2}}\lceil\sqrt{2}\rceil\) (for odd \(r\)) captures the primary magnitude of \(\sqrt{M}\), while the residue \(y = M - 2^r\) refines the estimate within the interval \([2^r, 2^{r+1})\). The factors \(\left(\frac{1}{2}-\frac{1}{16}\right)\) and \(\left(\frac{1}{2}+\frac{1}{16}\right)\) adjust the correction slope for even and odd \(r\), respectively, compensating for linearization errors and enabling partial error cancellation. Since the formulation relies primarily on shift-and-add operations, it provides an efficient piecewise-linear approximation that closely follows the square-root characteristic while ensuring low hardware complexity and suitability for FPGA-oriented implementations.
\subsection{Architectural Design Flow of the Proposed Square Rooter}
\begin{figure}[!t]
\hspace{-0.3cm}
\includegraphics[scale=0.55]{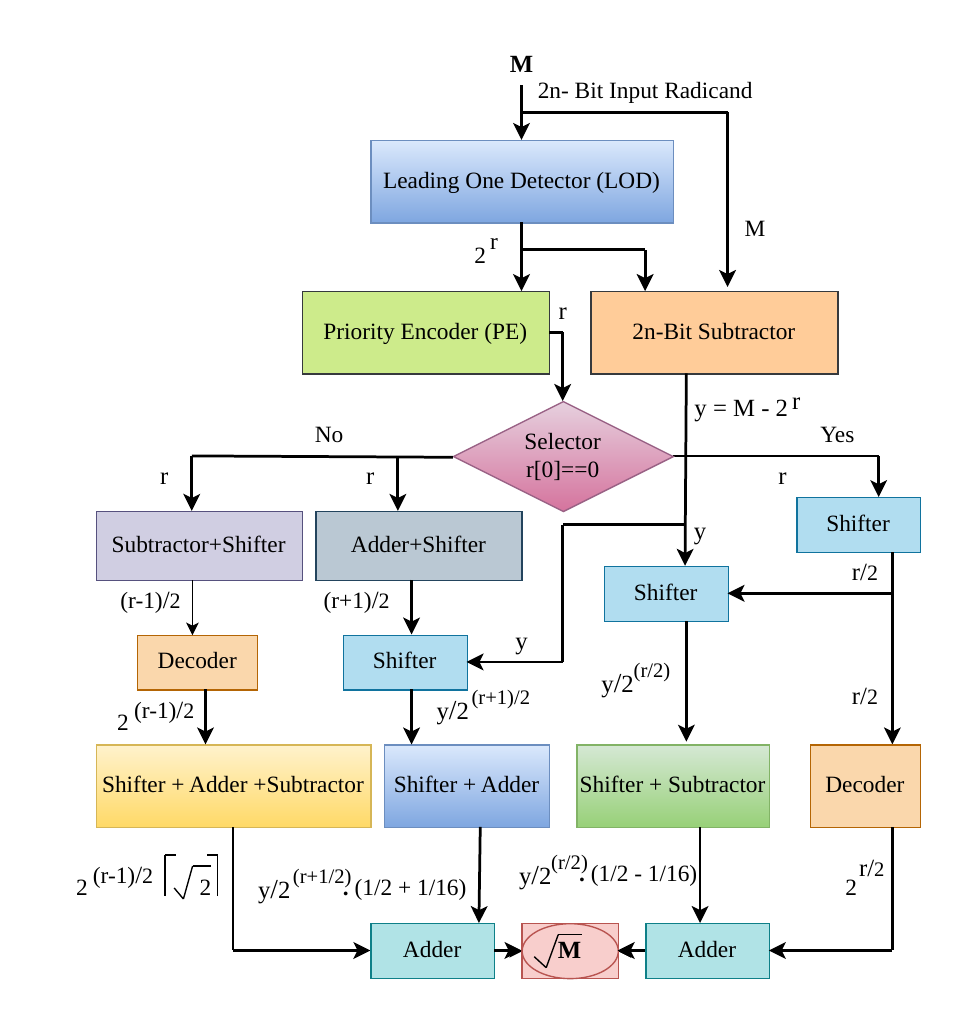}
\caption{Architectural Design Flow of the Proposed Square Rooter (EOSQR).}
  \label{fig:EOSQR}
\end{figure}

The detailed hardware architectural design flow of the proposed EOSQR design is illustrated in Fig.~\ref{fig:EOSQR}. The input radicand $M$ is first processed by the Leading One Detector (LOD) and Priority Encoder (PE) to determine the position of the leading one $r$, while a subtractor computes the residual term $y = M - 2^r$. A selector block determines the parity of $r$ by checking its least significant bit ($r[0]$) and accordingly activates the even or odd computation path.

The proposed design primarily employs simple combinational hardware units, such as shifters, adders, subtractors, and decoders, to compute the square root approximation. By avoiding multipliers and other complex arithmetic units and instead utilizing shift-based scaling with lightweight combinational logic, the proposed design achieves an energy-efficient implementation while maintaining high computational accuracy.
The proposed EOSQR design eliminates iterative operations and achieves an efficient balance between accuracy and energy efficiency, while maintaining comparable hardware complexity by leveraging a piecewise linear approximation with hardware-friendly coefficients.

An illustrative example consistent with Fig.~\ref{fig:EOSQR} and \eqref{eq:final} is presented in Table~\ref{tab:example}. In this case, the input radicand is \(32767\), whose exact square root is \(181\). The proposed design produces an output of \(184\), resulting in an error distance of only \(3\). This small deviation demonstrates the high accuracy of the proposed square-root computation.

\begin{table}[!t]
\raggedright
    \caption{Illustrative Example.}
    \label{tab:example}
    \begin{tabular}{|m{2.02cm}| m{1.53cm}| m{2.85cm}| m{0.75cm}|}
    \hline
        Operand & Block & Block Description  & Decimal\\
        \hline
       $M$ & Number & Input Radicand  & \bf $32767$ \\
        \hline
         $2^{r}$ & LOD & Largest power of $2$ & $16384$ \\
         $r$ & PE & Power of $2$ value  & $14$ \\
        $y$ & Subtractor & $y=M-2^{r}$  & $16383$ \\
         \hline
         $r[0]$ & Selector & $r[0]==0 \hspace{2mm}$(r=Even)  & \large - \\
        \hline
         $(\frac{r}{2})$ & Shifter & Right shifting $r$  & $7$ \\
       
        $(y/{2^{\frac{r}{2}}})$ & Shifter & Right shift $y$ by $(\frac{r}{2})$ & $127$ \\
         
        $(y/{2^{\frac{r}{2}}}).(\tfrac{1}{2}-\tfrac{1}{16})$ &Approximation & Using Shifter and Subtractor for $r=$even & $56$ \\

        $(2^{\frac{r}{2}})$ & Decoder & Raising $r/2$ result to the power of $2$  & $128$ \\
        \hline
        $\sqrt{M}$ & Adder & Final Addition ($128+56$)  & \bf $184$ \\
        \hline
    \end{tabular}
  
\end{table}
\section{Results and Discussion}
This section presents graphical and quantitative evaluations of the proposed Error Optimized Unsigned Square Rooter (EOSQR) for energy efficiency and computational accuracy. EOSQR is compared with the accurate baseline (ERAS) and state-of-the-art approximate square root designs across different approximation parameters ($t$). Furthermore, the Composite Efficiency Metric (CEM) is used to analyze the trade-off between accuracy preservation and hardware cost reduction.

\subsection{Approximation Parameter (t) Selection Strategy Across Compared Square Root Designs}

In AXSR3 \cite{SQ1}, the approximation parameter $t$ denotes the number of columns in which exact subtractor cells are replaced with approximate cells according to the triangular replacement (TR) scheme. Similarly, in MAHSQR \cite{SQ3}, the parameter $t$ represents the number of least significant bits assigned to the approximate computation block. In MAHSQR, increasing $t$ generally improves computational accuracy but increases hardware overhead (area, power, and delay), whereas reducing $t$ lowers hardware cost at the expense of accuracy. For a fair comparison, the approximation parameter $t$ for each design is selected to achieve comparable error levels across all square root designs, enabling a balanced evaluation of hardware overhead, energy efficiency, and computational accuracy. For LESQ-EC \cite{SQ2}, OLSR \cite{OLSR}, TSOSQR \cite{TSOSQR}, and the proposed EOSQR, the designs are inherently approximate and do not utilize a tunable approximation parameter $t$.

\begin{table*} [ht!]  
    \centering
    \begin{center}
     \caption{\small{Comparative evaluation of 16-bit unsigned square rooters with performance metrics across different designs.}} 
     \label{tab:comparison}
     \newcolumntype{C}[1]{>{\centering\arraybackslash}m{#1}}
   \begin{tabular}{ |m{1.8cm}|C{0.3cm}| C{0.6cm}| C{0.7cm}| C{0.8cm}| C{0.9cm} |C{0.9cm}| C{0.9cm} | C{0.85cm}|C{0.5cm}| C{0.8cm}|C{0.9cm}| C{0.9cm}| C{0.9cm}|} 
    \hline
        SQR  & t & \centering LUTs & DP & CPD  & PDP  & NMED & MRED & MED & ED & MSE & Power Savings & Resource Savings &  Latency Savings\\
        Designs  &  &  & (mW) & (ns)  & (pJ)  & ($\times 10^{-2}$)  & ($\times10^{-2}$) & & (max) & & $(\%)$ & $(\%)$ & $(\%)$ \\
        \hline
         \vspace{2.5mm}
        ERAS\cite{AC1} & - & $84$ & $18.88$ & $9.286$ & $175.321$ & \centering \large - & \centering \large -  &\centering \large -  &\centering \large -  &\centering \large -  &\centering \large -  &\centering \large -  & \large -\\
        \hline
         \vspace{2.5mm}
               
         AXSR3\cite{SQ1} &  $10$ & $61$ &  $12.54$ & $8.540$ & $107.092$ & $0.4822$ & $1.5461$  & $1.2291$ & \centering $14$ & $3.243$ & $33.58$ & $27.38$ & $8.03$ \\\vspace{2.5mm}
                
        & \centering $12$ & $55$ & $11.21$ & $7.704$ & $86.362$ & $1.6861$ & $4.7310$ & $4.3019$  & \centering $30$ & $32.68$ & $40.63$ & $34.52$ & $17.04$ \\
        \hline
         \vspace{2.5mm}
         LESQ-EC\cite{SQ2} & \centering - & $24$ & $3.12$ & $4.389$ & $13.694$ & $2.6819$ & $4.0257$ & $6.7822$  &\centering $21$ & $75.459$ & $83.47$ &$71.42$ & $52.73$ \\
      
                \hline
         \vspace{2.5mm}
                
         MAHSQR \cite{SQ3}  &\centering$6$ &  $34$ &  $4.24$ &  $4.680$ &  $19.843$ & $0.9849$ & $1.9158$ 
               & $2.5096$ & \centering $7$ &  $9.599$ &  $77.54$ &$59.52$ & $49.60$\\
               
        &\centering $4$ & $62$ &  $4.48$ & $ 4.997$ &  $22.387$ & $0.4819$ & $0.9417$
              & $1.2217$  & \centering $3$ & $2.389$ & $76.27$ &$26.19$ & $46.18$\\  
               \hline
         \vspace{1.5mm}
                 OLSR \cite{OLSR}  &\centering  - &  $22$ &  $3.24$ &  $4.345$ &  $14.078$ & $1.0889$ & $1.7798$ & $2.7798$
               & \centering $11$ & $15.879$ & $82.84$ &  $73.81$ &  $53.21$ \\  
              
                \hline 
                 \vspace{1.5mm}
            TSOSQR \cite{TSOSQR} &\centering  - &  $17$ &  $3.12$ &  $4.040$ &  $12.120$ & $1.1850$ & $1.9120$ &$3.0480$
               & \centering $11$ & $18.72$ & $83.47$  &  $79.76$ &  $56.49$ \\
               
                   \hline 
                 \vspace{1.5mm}
                
        \bf EOSQR  &\centering  - &  $32$ &  $4.24$ &  $4.356$ &  $18.469$ & $\mathbf{0.4741}$ & $\mathbf{0.7447}$ 
               & $\mathbf{1.2091}$ & $\mathbf{3}$ &  $\mathbf{2.279}$ &  $77.54$ &$61.91$ & $53.11$\\
                \hline             
    \end{tabular}
     \end{center}
\end{table*}
\subsection{Graphical Analysis: Behavior of Approximation}
 \begin{figure}[ht!]
		\centering
        \includegraphics[width=0.5\textwidth]{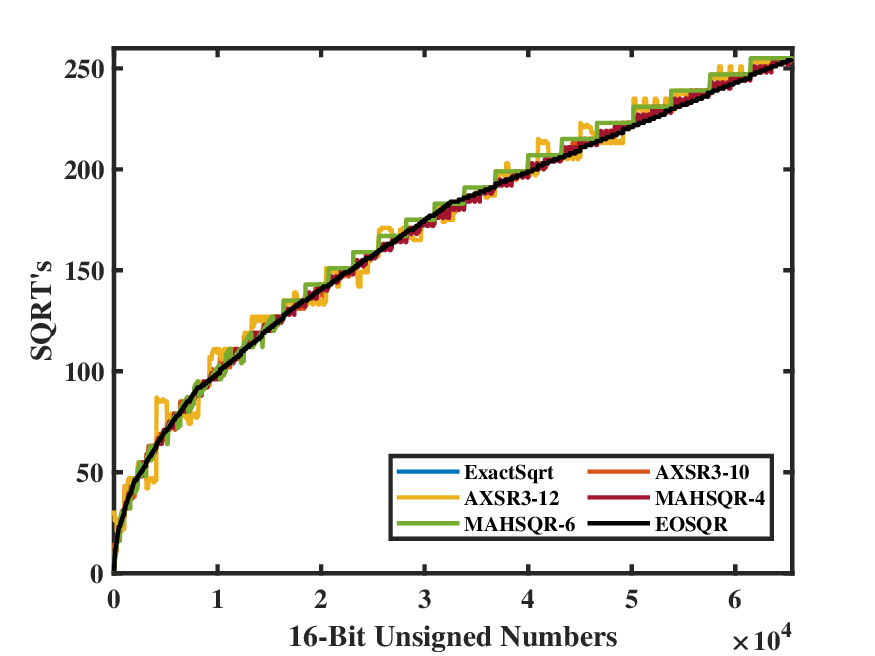}	
		\caption{Comparative graphical analysis of EOSQR with different approximate unsigned square rooters.}
		\label{fig:graph}
	\end{figure} 
Graphical comparisons provide an effective way to evaluate the accuracy and stability of approximate square root designs across the entire input range. For 16-bit inputs ($1$ to $65535$), a detailed comparison is performed against the exact square root, including the restoring-array and hybrid approaches, as well as the proposed EOSQR. Series-expansion-based square rooters are excluded from the graphical analysis, as they intentionally allow larger deviations to achieve higher hardware efficiency. The graphical results in Fig.~\ref{fig:graph} show that EOSQR closely follows the exact square root curve with minimal deviation, outperforming existing methods. In contrast, restoring-array and hybrid designs exhibit noticeable deviations due to scalability limitations and resource overhead. EOSQR, however, demonstrates near-exact behavior while maintaining lower hardware complexity, reduced resource utilization, and improved power–delay product (PDP), highlighting a favorable accuracy–efficiency trade-off for accuracy-aware and energy-constrained systems, discussed next.

\subsection{Comparative FPGA Implementation Results with Accuracy and Precision Evaluation}
All square root designs were described in Verilog HDL and synthesized using the Xilinx Vivado $2019.2$ design environment, targeting a Xilinx Artix-$7$ FPGA (XC7A35T-CPG236-1), while maintaining identical synthesis conditions for all designs to ensure a fair comparison. Hardware resource utilization was quantified using the Look-Up Table (LUT) count, which serves as a direct indicator of area complexity on the FPGA fabric. To obtain realistic dynamic power (DP) estimates, post-implementation timing simulations were performed using randomly generated input vectors, and the corresponding Switching Activity Interchange Format (SAIF) files were extracted. These SAIF files enable accurate modeling of node-level switching activity, allowing reliable estimation of dynamic power consumption. The critical path delay (CPD) for each design was derived from static timing analysis and represents the maximum propagation delay under worst-case operating conditions, while overall energy efficiency is measured using the power–delay product (PDP).

For energy-constrained signal and image processing applications, a $16$-bit unsigned square rooter producing an $8$-bit output provides an effective balance between implementation efficiency and computational precision. Table~\ref{tab:comparison} presents a comparative FPGA-based evaluation of state-of-the-art exact and approximate $16$-bit square root designs. In Table~\ref{tab:comparison} numerical accuracy is evaluated using standard error metrics, including Normalized Mean Error Distance (NMED), Mean Relative Error Distance (MRED), Mean Error Distance (MED), Maximum Error Distance (ED$_{\max}$), and Mean Squared Error (MSE), obtained through exhaustive MATLAB–HDL co-simulation over all possible $16$-bit unsigned input values.

A detailed comparison of the proposed Error-Optimized Unsigned Square Rooter (EOSQR) with existing square rooters demonstrates a well-balanced trade-off between accuracy, hardware utilization, and energy efficiency. EOSQR achieves the lowest error metrics NMED, MRED, MED, ED$_{\max}$, and MSE among the evaluated approximate designs, indicating superior numerical accuracy. Despite maintaining this high accuracy, EOSQR significantly improves hardware efficiency compared with the exact ERAS implementation, achieving a 77.54\% reduction in dynamic power, a 61.91\% reduction in LUT utilization, and a 53.11\% reduction in latency, resulting in a substantially lower PDP. Although AXSR3-10 \cite{SQ1} and MAHSQR-4 \cite{SQ3} exhibit comparable accuracy, they achieve this performance at considerably higher hardware overhead and PDP.

Series-expansion-based square rooters, including LESQ-EC \cite{SQ2}, OLSR \cite{OLSR}, and TSOSQR \cite{TSOSQR}, achieve high hardware efficiency by aggressively simplifying the computation, thereby reducing LUT utilization, dynamic power, and latency. However, this efficiency comes at the cost of degraded numerical accuracy. As shown in Table~\ref{tab:comparison}, these designs exhibit relatively higher error metrics, with NMED ranging from $1.08\times10^{-2}$ to $2.68\times10^{-2}$ and MED values exceeding $2.7$, indicating larger deviations from the exact square root. In contrast, the proposed Error-Optimized Square Rooter (EOSQR) is specifically designed to minimize approximation error while maintaining competitive hardware utilization with energy efficiency. EOSQR achieves significantly lower error metrics (NMED = $0.4741\times10^{-2}$, MRED = $0.7447\times10^{-2}$, MED = $1.2091$, and ED$_{\max}=3$), while retaining timing and energy characteristics comparable to series-expansion-based designs.

\subsection{Trade-off Evaluations using Composite Efficiency Metric (CEM)}
To ensure an equitable and quantitative comparison among various square rooters, a Composite Efficiency Metric (CEM) is formulated that integrates accuracy, energy efficiency, and hardware cost considerations \cite{TSOSQR}. It is mathematically expressed as:
\[
\text{CEM} = \frac{10^{-9} \times \text{ACM}}{\text{HWM}} 
= \frac{10^{-9} / (\text{MRED} \times \text{ED}_{\max})}{(\text{LUTs} \times \text{PDP})}
\]
Here, the \textit{ACM} (Accuracy Metric) encapsulates the error behavior, while \textit{HWM} (Hardware Metric) represents the implementation cost in terms of logic resource utilization and energy efficiency, as reflected in power-delay characteristics. The normalization factor of $10^{-9}$ is included to maintain numerical consistency across different metric scales.

\begin{figure}[ht!]
		\centering
        \includegraphics[width=0.5\textwidth]{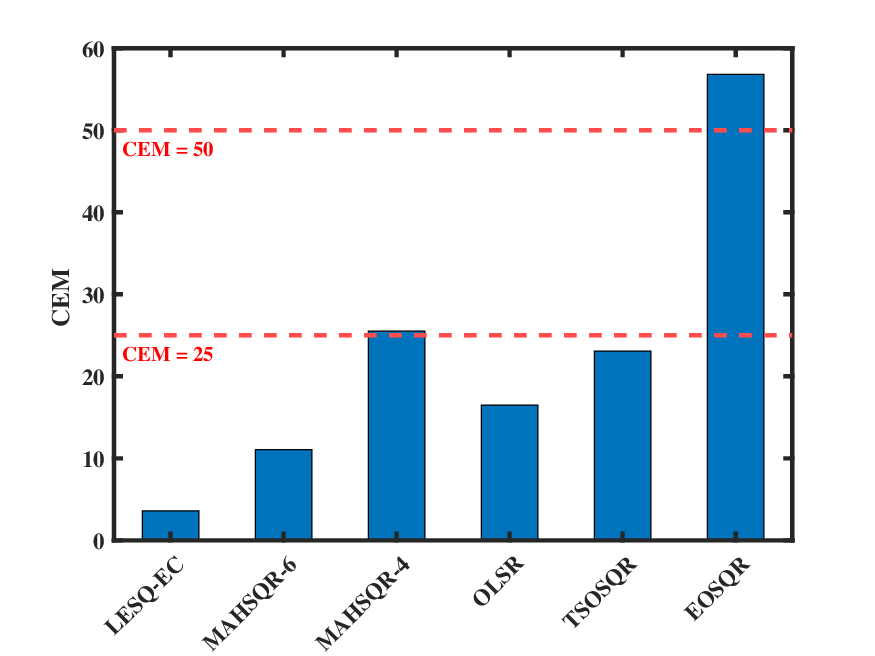}	
		\caption{Composite Efficiency Metric (CEM) assessment highlighting the proposed EOSQR’s dominance in accuracy preservation with an energy-efficient footprint for different approximate square rooters.}
		\label{fig: cem}
	\end{figure}

The proposed CEM framework provides a unified means to evaluate design quality by considering both average-case and worst-case performance. The Mean Relative Error (MRED) quantifies overall accuracy and sensitivity to functional nonlinearity, making it a meaningful indicator of expected operational precision in error-tolerant applications. Meanwhile, the Maximum Error Deviation (ED$_{\max}$) captures extreme deviations, ensuring reliability under worst-case scenarios.
On the hardware side, the Look-Up Table (LUT) count reflects spatial resource utilization, while the Power-Delay Product (PDP) represents the trade-off between dynamic power and latency, thereby indicating overall energy efficiency. By incorporating these four complementary dimensions, MRED, ED$_{\max}$, LUTs, and PDP, the CEM provides a holistic and application-relevant benchmark for evaluating approximate square rooters.

A higher CEM value indicates a design that delivers strong accuracy (both average and worst-case) with minimal resource usage and timing overhead. Thus, CEM serves as an effective figure-of-merit for identifying designs that optimally balance computational precision and hardware efficiency, which is essential for high-performance digital signal and image processing as well as accuracy-aware machine learning applications. Although restoring array-based designs (AXSR3) achieves lower computational errors, their high resource requirements significantly reduce overall efficiency, resulting in lower CEM scores. As shown in Fig.~\ref{fig: cem}, the proposed EOSQR design achieves the highest CEM value among all optimized designs.

These results demonstrate that EOSQR achieves high numerical accuracy while achieving substantial energy savings. By jointly optimizing power consumption, latency, and resource utilization, EOSQR achieves a superior CEM-based 
accuracy-efficiency trade-off compared with existing $16$-bit square rooters. This makes the proposed design well-suited for real-time signal and image processing applications on both FPGA and ASIC platforms.


\begin{table*}[ht!]
\centering 
		\caption{Comparison of PSNR ($dB$) and SSIM Metrics in Edge Detection Application over Diverse Image Sets for Different Square Rooters.}
		\label{tab:psnr}
		\vspace{2mm}
		
\begin{tabular}{|m{1.75cm} |p{0.3cm}|p{0.6cm}|p{0.6cm}|p{0.6cm}|p{0.6cm}|p{0.6cm}|p{0.6cm}|p{0.6cm}|p{0.6cm}|p{0.6cm}|p{0.6cm}|p{0.6cm}|p{0.6cm}|p{0.6cm}|p{0.6cm}|}
\hline
\centering\textbf{SQR} & \centering\textbf{t} & \multicolumn{2}{|p{1.2cm}|}{\centering\textbf{Pirates}} & \multicolumn{2}{|p{1.2cm}|}{\centering\textbf{Cameraman}} & \multicolumn{2}{|p{1.2cm}|}{\centering\textbf{Barbara}} & \multicolumn{2}{|p{1.2cm}|}{\centering\textbf{House}} & \multicolumn{2}{|p{1.2cm}|}{\centering\textbf{Peppers}} & \multicolumn{2}{|p{1.2cm}|}{\centering\textbf{Mug}} & \multicolumn{2}{|p{1.2cm}|}{\centering\textbf{Average}}  \\ \hline

\centering\textbf{Designs} & {\textbf{}} &  {\textbf\small{PSNR}} & {\textbf\small{SSIM}} & {\textbf\small{PSNR}} & {\textbf\small{SSIM}} &  {\textbf\small{PSNR}} & {\textbf\small{SSIM}} & {\textbf\small{PSNR}} & {\textbf\small{SSIM}} &  {\textbf\small{PSNR}} & {\textbf\small{SSIM}} & {\textbf\small{PSNR}} & {\textbf\small{SSIM}} & {\textbf\small{PSNR}} & {\textbf\small{SSIM}}   \\ \hline

\multirow{2}{*}{AXSR3\cite{SQ1}}& \centering $12$ & $26.36$ & $0.843$ & $24.19$ & $0.928$ & $27.14$ & $0.876$ & $24.43$ & $0.916$ & $26.83$ & $0.879$ & $26.99$ & $0.919$ & $25.99$ & $0.894$  \\ 
  & \centering $10$ & $34.71$ & $0.984$ & $33.27$ & $0.983$ & $35.92$ & $0.981$ & $33.01$ & $0.982$ & $35.45$ & $0.984$ & $36.05$ & $0.981$ & $34.74$ & $0.982$\\ 
			\hline
LESQ-EC\cite{SQ2} & \centering - & $36.76$ & $0.970$ & $38.29$ & $0.977$ & $35.32$ & $0.962$ & $38.46$ & $0.976$ & $37.37$ & $0.970$ & $38.67$ & $0.965$ & $37.48$ & $0.970$  \\ 
			\hline

            \multirow{2}{*}{\centering MAHSQR\cite{SQ3}} 
& \centering $6$ & $39.17$ & $0.942$ & $40.66$ & $0.953$ & $39.21$ & $0.926$ & $40.21$ & $0.951$ & $39.24$ & $0.942$ & $40.94$ & $0.949$ & $39.91$ & $0.944$  \\

& \centering $4$ & $44.66$ & $0.965$ & $45.04$ & $0.967$ & $44.69$ & $0.954$ & $44.92$ & $0.973$ & $44.62$ & $0.965$ & $44.87$ & $0.969$ & $44.80$ & $0.966$   \\

			\hline
             OLSR\cite{OLSR} & \centering - & $42.51$ & $0.959$ & $44.17$ & $0.979$ & $41.33$ & $0.965$ & $45.65$ & $0.978$ & $44.10$ & $0.973$ & $43.13$ & $0.972$ & $43.48$ & $0.971$  \\ 
			
			\hline
            TSOSQR\cite{TSOSQR} & \centering - & $42.22$ & $0.956$ & $43.88$ & $0.978$ & $40.93$ & $0.962$ & $43.83$ & $0.977$ & $42.87$ & $0.971$ & $43.99$ & $0.976$ & $42.95$ & $0.971$  \\ 
			
			\hline
			\bf EOSQR  & \centering -  & $\mathbf{49.61}$ & $\mathbf{0.985}$ & $\mathbf{50.58}$ & $\mathbf{0.984}$ & $\mathbf{48.79}$ & $\mathbf{0.982}$ & $\mathbf{50.64}$ & $\mathbf{0.982}$ & $\mathbf{49.73}$ & $\mathbf{0.985}$ & $\mathbf{51.79}$ & $\mathbf{0.983}$ & $\mathbf{50.19}$ & {$\mathbf{0.984}$ } \\ 
			\hline
			
		\end{tabular}
\end{table*}

\begin{figure*}[ht!]
	\centering
    \subfloat[\scriptsize Original]{\includegraphics[scale=0.53]{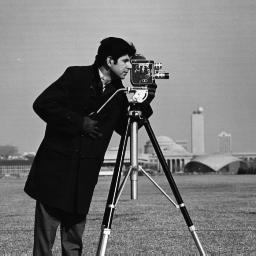}}
    \hfill
    \subfloat[\scriptsize ERAS]{\includegraphics[scale=0.53]{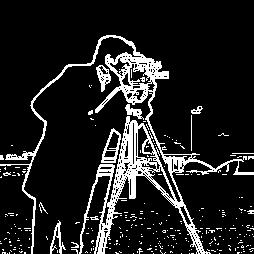}}
    \hfill
	\subfloat[\scriptsize AXSR3-12]{\includegraphics[scale=0.53]{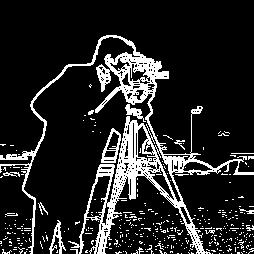}}
    \hfill
	\subfloat[\scriptsize AXSR3-10]{\includegraphics[scale=0.53]{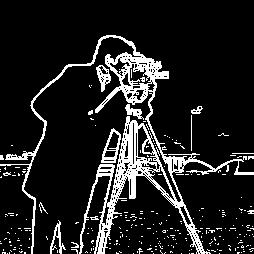}}
    \hfill
	\subfloat[\scriptsize LESQ-EC]{\includegraphics[scale=0.53]{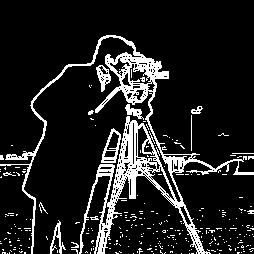}}
      \hfill
	\subfloat[\scriptsize MAHSQR-6]{\includegraphics[scale=0.53]{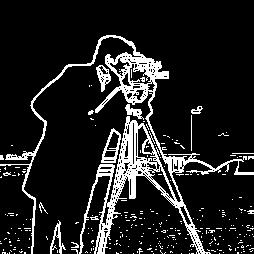}}
      \hfill
	\subfloat[\scriptsize MAHSQR-4]{\includegraphics[scale=0.53]{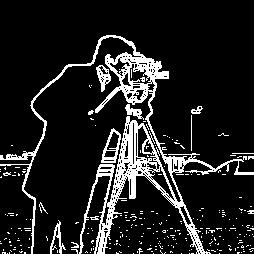}}
      \hfill
	\subfloat[\scriptsize OLSR]{\includegraphics[scale=0.53]{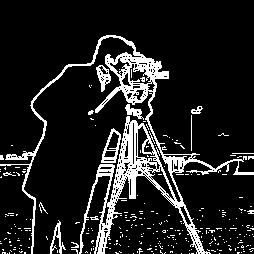}}
    \hfill
	\subfloat[\scriptsize TSOSQR]{\includegraphics[scale=0.53]{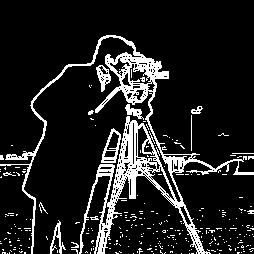}}
    \hfill
	\subfloat[\scriptsize EOSQR]{\includegraphics[scale=0.53]{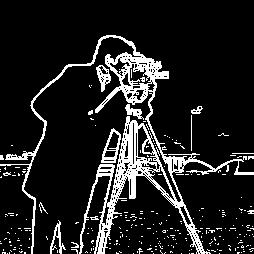}}	
      \hfill
	\caption{Visual results of Sobel edge detection using multiple approximate square rooters, parameterized by the approximation factor $t$.}
\label{fig:edge_detection_results}
\end{figure*}

\section{Applications}
To demonstrate the practicality and robustness of the proposed Error Optimized Unsigned Square Rooter (EOSQR), its performance has been validated across diverse application domains, including image processing and machine learning. The EOSQR achieves a remarkable balance between computational accuracy and energy efficiency, delivering significantly lower error metrics while maintaining comparable hardware utilization. To substantiate the design’s efficiency under real-world workloads, it has been employed in several representative applications: (i) Sobel edge detection, to evaluate its sensitivity to pixel intensity gradients in image boundary extraction; (ii) K-Means clustering for color quantization, serving as an unsupervised machine learning benchmark that leverages the square root operation in Euclidean distance computation; and (iii) K-Nearest Neighbor (KNN) classification, representing a supervised learning scenario where classification accuracy under approximate arithmetic is examined. These applications confirm the design’s ability to maintain high accuracy and visual fidelity, demonstrating its suitability for energy-efficient embedded and FPGA-based systems.

\subsection{Edge Detection}
Due to the inherent constraints of human visual perception, approximate computing has become increasingly prevalent in image analysis and computer vision, where a tolerable margin of inaccuracy can be traded for significant gains in efficiency~\cite{FN4}. In this application, the performance of the proposed square rooter is evaluated against an exact edge-detection implementation. The Sobel operator identifies intensity transitions in an image by convolving pixel neighborhoods with predefined kernels to estimate the horizontal ($G_x$) and vertical ($G_y$) gradient components. The overall gradient magnitude ($G$) is obtained as $G = \sqrt{ (G_{x})^2 + (G_{y})^2}$, which quantifies the edge strength across both directions. In this work, both accurate and approximate 16-bit square root designs are employed to compute $G$ within the Sobel framework. The experimental study is performed in MATLAB–HDL co-simulation using Simulink, where the Verilog-based square rooter is integrated.

Table~\ref{tab:psnr} presents a comparative analysis of various square rooters, where the Peak Signal-to-Noise Ratio (PSNR) and Structural Similarity Index Metric (SSIM) are evaluated using the exact square root output as the reference benchmark for edge detection. The assessment is conducted on six $8$-bit grayscale images, namely Pirates, Cameraman, Barbara, House, Peppers, and Mug, each with different spatial resolutions. The results clearly demonstrate that the proposed EOSQR design achieves the highest average PSNR ({50.19~dB}) and average SSIM ({0.984}) among all competing methods, exhibiting stable and reliable performance and highlighting its superior capability to preserve image quality. These results affirm that the proposed design achieves superior accuracy while preserving strong energy efficiency.
To visually demonstrate the edge detection process, the “\textit{Cameraman}" image is presented in Fig. \ref{fig:edge_detection_results}, together with the corresponding edge-detected outputs obtained using different square rooters for comparative analysis.

\subsection{K-Means Clustering as a Machine Learning Benchmark for Color Quantization}

To demonstrate the applicability of the proposed approximate square rooter in machine learning and image compression, its integration into the K-Means clustering framework for color quantization is presented. K-Means clustering is a well-established unsupervised learning algorithm extensively employed in image processing to reduce the color space by grouping similar pixel intensities into representative clusters. This process effectively decreases the number of unique colors in an image, thereby enabling efficient compression while maintaining acceptable perceptual quality. Such characteristics make the technique particularly suitable for memory-constrained and bandwidth-limited environments. The algorithm partitions the RGB color space into $K$ clusters and iteratively refines their centroids to minimize intra-cluster variance. During each iteration, the Euclidean distance between pixel vectors and cluster centroids is computed to update memberships and recompute centroid positions. This iterative distance evaluation serves as a rigorous benchmark for assessing the computational accuracy and efficiency of the proposed square rooter.

\begin{figure}[ht]
\centering

\subfloat[\scriptsize Original]
{\includegraphics[scale=0.248]{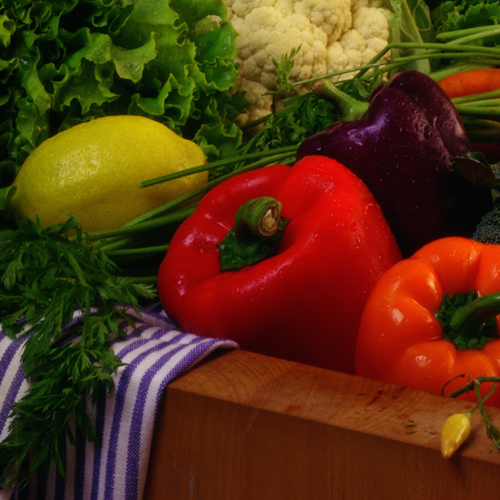}}
\hfill
\subfloat[\shortstack{\scriptsize AXSR3-10\\
\scriptsize (PSNR = 28.34)\quad (SSIM = 0.831)}]
{\includegraphics[scale=0.248]{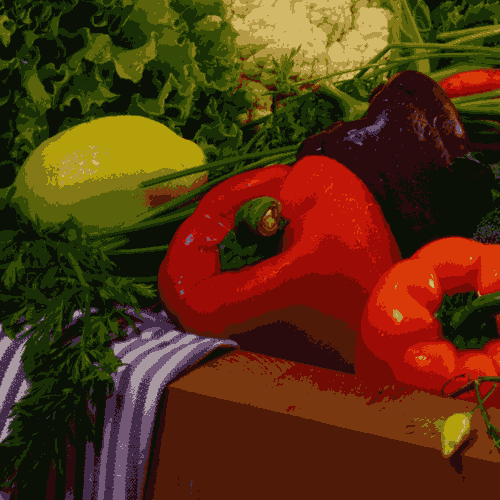}}
\hfill

\par\medskip

\subfloat[\shortstack{\scriptsize MAHSQR-4\\
\scriptsize (PSNR = 29.94)\quad (SSIM = 0.857)}]
{\includegraphics[scale=0.248]{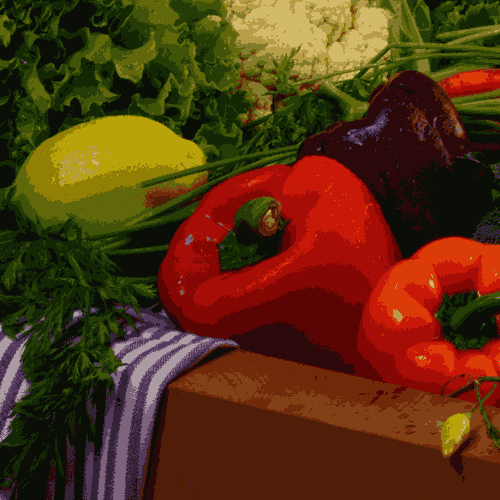}}
\hfill
\subfloat[\shortstack{\scriptsize EOSQR\\
\scriptsize (PSNR = 30.21)\quad (SSIM = 0.862)}]
{\includegraphics[scale=0.248]{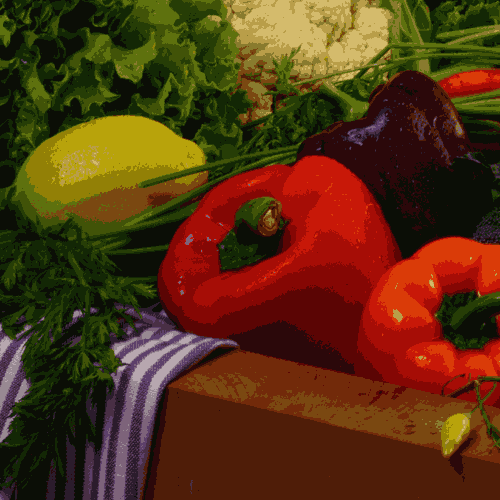}}

\caption{Visual representation of K-means clustering for color quantization.}
\label{fig:color_quantization_results}
\end{figure}
The K-Means clustering algorithm was implemented in Python using the PyCharm $2025.1.3$ development environment. Based on the accuracy optimization criteria  (\text{MRED} $\leq 1.5\%$ and \text{MSE} $\leq 3.5$) from Table~\ref{tab:comparison}, the corresponding approximate square rooters were individually modeled and integrated into the clustering process for comprehensive performance evaluation. A standard benchmark image, \textit{``Peppers''}, was quantized into $25$ representative colors over $5$ iterations, and reconstruction quality was evaluated using PSNR and SSIM. To visually assess the impact of different square rooters on clustering performance, the original image and its color-quantized outputs generated using various approximate square rooter implementations are presented in Fig.~\ref{fig:color_quantization_results}, along with the corresponding evaluation metrics. The proposed EOSQR achieved the highest PSNR of $30.21$\,dB and SSIM of $0.862$, producing the best visual quality. These results confirm the effectiveness of EOSQR for machine-learning-based image compression, offering high computational accuracy while improving hardware and energy efficiency compared with error-optimized designs.
\subsection{K-Nearest Neighbor (KNN) Classification using Approximate Square Rooter}
To evaluate the practical impact of the proposed Error-Optimized Square Rooter (EOSQR) in another machine learning application, the K-Nearest Neighbor (KNN) classifier is implemented in Python using the MNIST image dataset. KNN is a widely adopted non-parametric supervised learning technique used in pattern recognition, image classification, and data mining applications~\cite{KNN1}, operating on the principle that similar data points reside in proximity within the feature space. In this algorithm, classification is performed by computing the Euclidean distance between a test sample and all training samples, selecting the $k$ nearest neighbors, and assigning the majority class label. Since KNN relies heavily on intensive distance computations, the square root operation becomes a significant contributor to computational complexity. To address this, the exact square root is replaced with the proposed $16$-bit unsigned EOSQR within the Euclidean distance computation stage. 
Furthermore, to ensure hardware consistency, all distance calculations use a $16$-bit constrained framework, where accumulated squared differences are limited to $16$ bits before square root evaluation. This enables direct assessment of how arithmetic-level approximation affects end-to-end machine learning performance.

\begin{table}[ht!]
\centering
\caption{Performance Comparison of KNN Classifier Using Different Square Rooters}
\label{tab:knn_comparison}
\renewcommand{\arraystretch}{1.0}
\setlength{\tabcolsep}{3pt}
\footnotesize
\newcolumntype{C}[1]{>{\centering\arraybackslash}m{#1}}
\begin{tabular}{|m{2cm}|C{1.2cm}|C{1cm}|C{1cm}|C{1cm}|C{1.2cm}|}
\hline
\centering\textbf{Square Rooter} &
\centering\textbf{Accuracy (\%)} &
\centering\textbf{MAE} &
\centering\textbf{MSE} &
\centering\textbf{MAXE} &
\centering\textbf{Accuracy Drop (\%)} \tabularnewline
\hline

ERAS \cite{AC1} 
& 8.7 
& 0 
& 0 
& 0 
& 0 \\

\hline

AXSR3-10 \cite{SQ1}
& 8.1 
& 2.46 
& 9.35 
& 7.70 
& 0.60 \\

\hline
MAHSQR-4 \cite{SQ3} 
& 8.3 
& 1.87 
& 3.82 
& 3.82 
& 0.40 \\

\hline
\textbf{EOSQR} 
& 8.6 
& 1.52 
& 3.22 
& 3.62 
& 0.10 \\

\hline
\end{tabular}
\end{table}
According to the previously adopted optimization criterion for error-optimized designs suitable for KNN classification, the classifier outputs standard performance metrics such as accuracy and precision, along with distance error metrics including Mean Absolute Error (MAE), Mean Squared Error (MSE), and Maximum Error (MAXE) to quantify approximation-induced distortion. As summarized in Table~\ref{tab:knn_comparison}, experimental results indicate that KNN is inherently tolerant to square root approximation, since classification depends mainly on the relative ordering of distances rather than their exact magnitudes. As long as the nearest-neighbor ranking remains unchanged, classification accuracy is largely preserved.
The proposed EOSQR achieves an accuracy of $8.6\%$, closely matching the exact implementation ($8.7\%$) while exhibiting lower error metrics than AXSR3-$10$ and MAHSQR-$4$, resulting in only a $0.10\%$ reduction in accuracy. These results show that comparable recognition performance is maintained despite arithmetic approximation, while achieving improved efficiency in terms of delay, power consumption, and logic utilization. This confirms the suitability of EOSQR for machine learning accelerators in low-power edge and embedded systems.
\section{Conclusion}

This paper presents a piecewise-linear approximation-based energy-efficient Error-Optimized Unsigned Square Rooter (EOSQR) that achieves high numerical accuracy with low hardware complexity, for accuracy-critical applications. 
Through analytical, graphical, and quantitative evaluations, EOSQR demonstrates superior accuracy among approximately $16$-bit square-rooters while preserving strong energy efficiency. Compared with the exact ERAS implementation, EOSQR achieves up to $77.54\%$ reduction in dynamic power, $61.91\%$ reduction in LUT utilization, and $53.11\%$ reduction in latency, resulting in a substantially lower power–delay product than designs with similar accuracy. FPGA synthesis on a Xilinx Artix-$7$ platform confirms the practical efficiency of the proposed architecture. Furthermore, integration into representative workloads, including edge detection, K-Means color quantization, and KNN classification, demonstrates that EOSQR maintains high visual fidelity and classification accuracy under arithmetic approximation. The proposed design also achieves a superior Composite Efficiency Metric (CEM), highlighting its effectiveness in jointly optimizing accuracy and energy efficiency. Consequently, EOSQR provides a promising solution for energy-efficient FPGA and ASIC implementations in real-time signal processing, image processing, and edge AI systems.


\section*{Acknowledgments}
We thank the Visvesvaraya PhD Scheme for Electronics and IT: Phase-II (Ref.no.PhD-02/2022/25), the Science and Engineering Research Board (SERB): MTR/2021/00841, the Indo-French Centre for the Promotion of Advanced Research (CEFIPRA), India: 7143-SARI, and the Indian Institute of Technology Goa (IIT Goa) for financial support. 

\bibliographystyle{IEEEtran}
\bibliography{casreference}

\vspace{-40pt}
\begin{IEEEbiography}[{\includegraphics[width=1in,height=1.25in,clip,keepaspectratio]{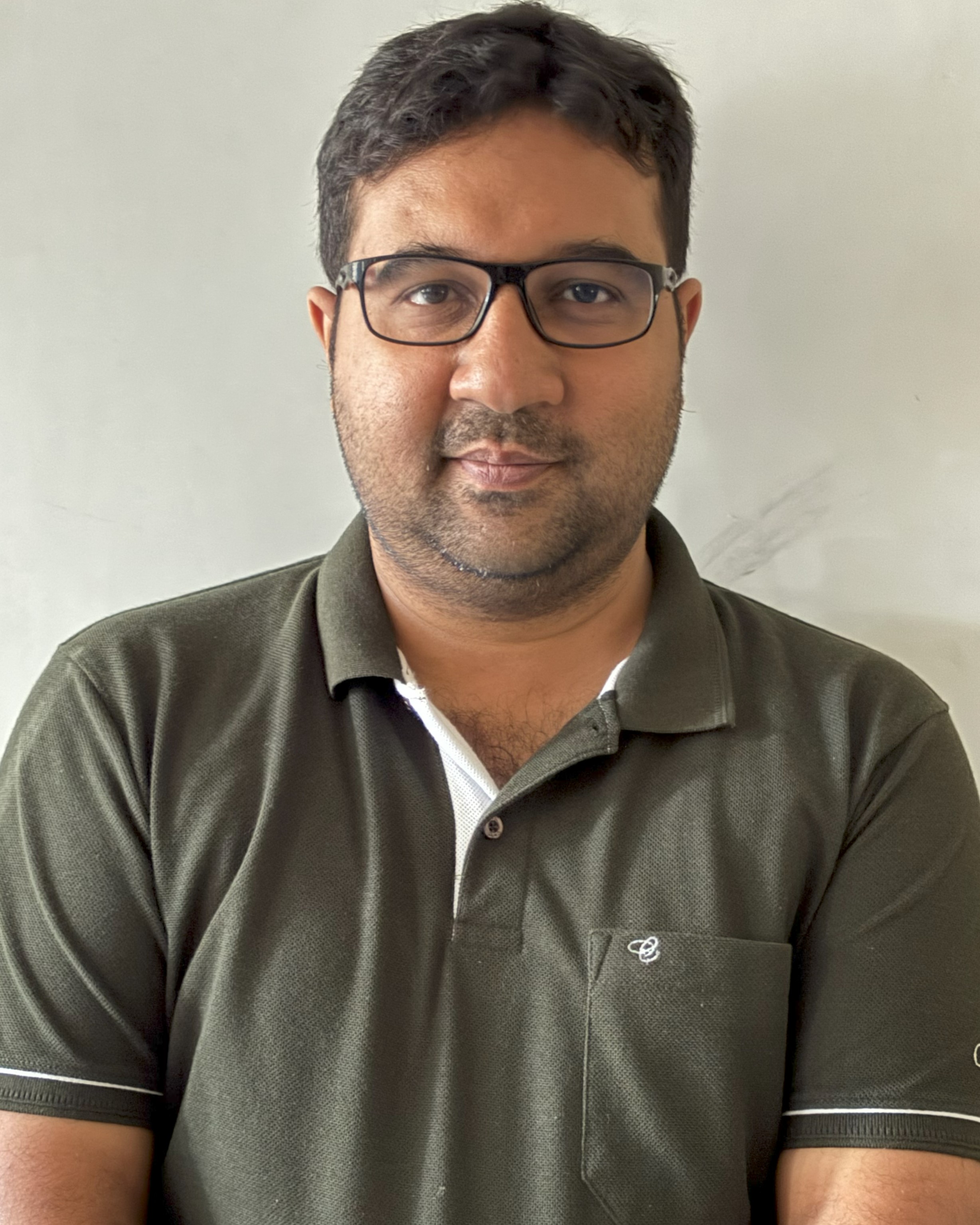}}]{Prateek Goyal}
received a B.Tech. degree in Electronics and Communication Engineering from
Dr.KNMIET Modinagar, UP, India, in 2010, a PG Diploma in Wireless Technology from CDAC, Noida, UP, India, in 2011, and a M.Tech. degree in Electronics and Communication Engineering from AMITY University, Noida, UP, India, in 2015. 
He is currently a Ph.D. Research Scholar in the School of Electrical Sciences at the Indian Institute of Technology Goa (IIT Goa), India. His research interests focus on low-power IC design and approximate computing.
\end{IEEEbiography}
\vspace{-40pt}
\begin{IEEEbiography}[{\includegraphics[width=1in,height=1.25in,clip,keepaspectratio]{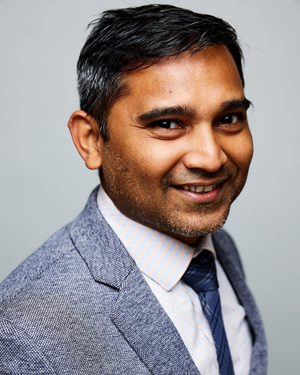}}]{Sujit Kumar Sahoo}
(SM’16, M’11) received a B.Tech. (Hons.) degree in Electrical Engineering in 2006 from the National Institute of Technology, Rourkela, India, and a Ph.D. in Electrical and Electronic Engineering in 2014 from the Nanyang Technological University, Singapore. From October 2006 to December 2007, he was a software engineer at Sasken Communication Technologies Ltd., Bangalore, India. From January 2012 to July 2018, he was a researcher at Nanyang Technological University, Singapore. From July 2018 to November 2023, he was an Assistant Professor at the School of Electrical Sciences at the Indian Institute of Technology Goa (IIT Goa) in India, where he presently serves as an Associate Professor. His research interests include sparse representation, compressed sensing, image/signal processing, computational imaging, inverse problems, and approximate computing.
\end{IEEEbiography}

\end{document}